\newcommand{\ignore}[1]{}
\ignore{}
\documentclass[11pt]{article}
\usepackage[xdvi]{epsfig}

\newcommand{\half}{ {\scriptstyle \frac{1}{2} } }
\newcommand{\quarter}{ {\scriptstyle \frac{1}{4} } }
\newcommand\be{\begin{equation}}
\newcommand\ee{\end{equation}}
\newcommand\bea{\begin{eqnarray}}
\newcommand\eea{\end{eqnarray}}
\newcommand\nn{ \nonumber\\}

\newcommand{\text}[1]{\qquad \mbox{#1} \qquad}

\title{Brane World Gravity  in an $AdS$ Black Hole
\thanks{Brown-HET-1316: This work was supported in part by the 
Department of Energy under Contracts No. DE-FG02-91ER40676 and No.
DE-FG02-91ER40688}}
\author{Richard  C. Brower
\\ Physics 
Department\\Boston University\\
Boston, MA 02215, USA \\
\and
Samir 
D. Mathur\\
Department of Physics\\
Ohio State University \\
Columbus 
OH 43210, USA \\
\and
Chung-I Tan \\
Physics Department \\
Brown 
University\\
Providence, RI 02912, USA}

\begin{document}

\maketitle
\vskip - 2cm
\begin{abstract}
We consider a model of brane world gravity in the context of
non-conformal non-SUSY matter.  In particular we modify the earlier
strong coupling solution to the glueball spectrum in an $AdS^7$ Black
Hole by introducing a Randall-Sundrum Planck brane as a UV
cut-off. The consequence is a new normalizable zero mass tensor state,
which gives rise to an effective Einstein-Hilbert theory of gravity,
with exponentially small corrections set by the mass gap to the
discrete glueball spectrum. However the simplest microscopic theory
for the Planck brane is found to have a tachyonic instability in the
radion mode.
\end{abstract}
\newpage

\section{Introduction}
 
An interesting idea that has emerged in recent years is that the
energy scale of gravity might not be significantly higher than the
scale of other fundamental interactions~\cite{Dimo, Dienes}. Models to
achieve this goal have invoked extra dimensions, which may be either
compact circles, or ``effectively compact'' directions where the
spacetime metric decreases exponentially with the proper distance in
the internal direction~\cite{RS1,RS2}. Gravity lives in all the
dimensions, while matter is confined to a hypersurface (the brane).
 
We study a model that follows the spirit of the latter class of
theories, but which has some interesting
features~\cite{maldacena,wittenO,gkp,wittenSuskind} not found in
earlier models. Our compact direction ends in a smooth way at some
value of the internal
coordinate~\cite{wittenT,csaki,jev,bmt1,bmt2,cm,oz,PolStras,StrasKleb},
instead of terminating in a second brane or continuing forever. Thus
the ``warped spacetime'' lies in the region $r_{min} <r<r_c$, with the
graviton localized near a brane at $r_c$ and $r=r_{min}$ being a
smooth end to the internal radial direction $r$.  Such a metric ending
smoothly at $r=r_{min}$ (and becoming approximately anti-de-Sitter at
large $r$) arises in the $AdS^7 \times S^4$ Black Hole gravity dual of
d=4 QCD at strong coupling~\cite{wittenT}.  We modify the metric
emerging from this construction, by placing a ``Planck brane'' at
$r=r_c$ and impose reflection symmetry about $r=r_c$.

The original AdS/CFT dual description for $QCD_4$ at strong coupling
suggested by Witten~\cite{wittenT} is in fact a finite temperature 5-d
Yang-Mills theory at the boundary of a deformed AdS space with radial
coordinate $r_{min}<r<\infty$.  Anti-periodic boundary condition on
the thermal axis for fermions breaks conformal and SUSY symmetries
giving rise to a discrete ``glueball'' spectrum with mass scale set by
$\Lambda_{qcd} = r_{min}/R^2_{ads}$ or equivalently by the temperature
of the resultant Black Hole.  In our present analysis we find the
following features.  The Planck brane at $r=r_c$ must be asymmetric -- it
must have one tension along the ``spacetime'' directions of the brane,
but a different tension along the ``thermal'' direction $\tau$, the
ratio of these two tensions fixing the brane location to  $r=r_c$. We
construct a model, which yields a brane with effective tensions that
are asymmetric in the desired manner.

We then look at the small fluctuations around this background. We find
a massless graviton ``localized on the brane''. But we also have a
``radion'' mode that is essentially a fluctuation of the proper
distance from the brane at $r_c$ to the horizon of the Black Hole at
$r=r_{min}$. For our microscopic construction of the asymmetric brane,
we find that this radion mode is an {\it unstable} 
solution. We then examine a 1-parameter family of ``effective
potentials'' for the brane, which generalize the behavior found in this
explicit brane construct and solve analytically for all zero-mass
excitations.  We find that beyond a certain critical value for this
parameter the radion mode indeed becomes stable. However we have not
been able to find an explicit construction of branes which leads to
these effective potentials.

An interesting feature of the model is the presence of both a graviton
mode localized at the Planck brane and a discrete set of radial ``KK
states'' analogous to the glueball modes found in the dual description
of QCD.  The large mass hierarchy required for the effective 4-d
Planck mass relative to the QCD scale is exponential in the proper
distance from $r_{min}$ to $r_c$. The ``glueball states'' suffer only
an exponentially small correction due to the fact that the coordinate
$r$ has an UV cut-off at $r=r_c$ instead of continuing to infinity. It
is true that the model we are considering has not strictly been
derived from string theory and therefore may not exhibit Gravity/Gauge
duality. But we have chosen the metric for $r < r_c$ to be equal to
that found in the strong coupling dual to d=4 QCD, so it is
interesting to speculate that these ``glueball modes'' still represent
some kind of gauge excitations in the current model as well. The fact
that both gravity and glueball modes are excitations of a single
string description in AdS space opens up interesting questions on the
relations between these two entities, and we comment on these issues
at the end of the paper.

\newpage
\section{The 
Model}
\label{sec:model}

We introduce an effective low energy model for brane world gravity
interacting with non-conformal matter. Our approach is based on
modifying the AdS/CFT example proposed by Witten~\cite{wittenT} as a
gravity dual for $QCD_4$ in the strong coupling limit. We modify the
UV behavior with the insertion of a so called ``Planck brane'' in the
spirit of Randall-Sundrum~\cite{RS1,RS2} brane world gravity.

Let us recapitulate Witten's suggestion of the gravity dual of
$QCD_4$.  One begins with the AdS/CFT correspondence for the 11-d
M-theory background metric $AdS^7
\times S^4$.  The 11th coordinate 
is compactified on a small circle (with radius $R_{11} = g_s l_s$)
reducing M-theory to IIA string theory and the boundary (0,2) 6-d CFT
to 5-d SUSY Yang-Mills.  Then a second circle is introduced with
anti-periodic boundary coordinate for all fermionic modes so that
conformal and all supersymmetries are broken.  The second circle will
be designated here by $\tau$ or simply the thermal coordinate.  It is
conjectured that at weak coupling the corresponding field theory is a
confining 4-d Yang-Mills theory.  This new metric is an AdS Black Hole
solution to the bosonic sector of 11-d supergravity,
\be S = - 
\frac{1}{2 \kappa_{11}} \int d^{11}x \sqrt{-g_{11}} \; ( R_{11}
   - 
|F_4|^2) + \frac{1}{12 \kappa_{11} }\int A_3 \wedge F_4 \wedge F_4
+ 
\mbox{fermions}  \; ,
\ee
written in term of the metric tensor $g_{MN}$ and the 3-form gauge
field $A_{MNL}$ and its field strength.  In the black brane solution a
constant background for $A_{MNL}$ for N units of magnetic flux gives
rise to an effective cosmological constant; ignoring fluctuations in
$A$ and nonzero R charges in $S^4$, it is adequate for our
present purpose to consider a simpler action in the $AdS^{D}$
subspace,
\be
S  =  - \frac{1}{2 
\kappa_D} \int_M   d^Dx \sqrt{-g} \; ( R -  2 \Lambda ) \; 
.
\label{eq:EH}
\ee
$M_D \sim 1/\kappa_D^{1/d}$ is the bulk Planck mass in $AdS^{d+2}$
with $D = d+2$.  The $D = d+2$ coordinates are designated by $x^M =
(r,\tau,x^\mu)$ with $\mu = 1,\cdots d$. We also find it convenient
to consider a general $AdS^{d+2}$ instead of restricting our discussion to
$AdS^7$ appropriate for the M theory construction.

\subsection{ Black Hole Background} 

Substituting in to Einstein's equations the ansatz
\be ds^2 = 
\hat g_{MN} dx^M dx^N \equiv \frac{1}{f(r)} dr^2 + f(r)
d\tau^2 + 
f_0(r) g_{\mu\nu} (x) dx^\mu dx^\nu \; , \ee
with one ``radial'' coordinate $r$, one angular coordinate $\tau$
periodic in $[0,\beta)$ and $f_0(r) \equiv k^2 r^2$, one finds the
general solution, $f(r,c) = c + (k r)^2 -
(kr_{min})^{d+1}/(kr)^{d-1}$, with $c = 0,-1,+1$ for $ds^2_d =
g_{\mu\nu} (x) dx^\mu dx^\nu$ being Minkowski (or Euclidean), de Sitter
and anti-de Sitter respectively.  These spaces are all asymptotic to
$AdS^{d+2}$ as $r \rightarrow \infty$ with the AdS radius $R_{ads} =
1/k$ fixed by the cosmological constant, $\Lambda = - (D-1)(D-2) k^2/2
\; .$ In particular we are interested in the asymptotically flat
Minkowski background $ds^2_d = \eta_{\mu\nu} dx^\mu dx^\nu$, thus  r $c=0$ and 
\be
f(r) \equiv (k r)^2 - 
\frac{(kr_{min})^{d+1}}{(kr)^{d-1}}
\ee
The period for the thermal axis ($\tau \rightarrow \tau + \beta$), or
inverse ``Hawking temperature", is fixed to be $\beta = (4
\pi)/((D-1) k^2 r_{min})$ by the requirement that the
horizon is a coordinate singularity at $r = r_{min}$. For the case
$r_{min}=0$, the space becomes pure AdS, $f(r) \rightarrow f_0(r) =
k^2 r^2$ and there is no horizon and no condition on the
periodicity. Often it is more convenient to work with units where
$R_{ads} \equiv k^{-1} =1$. We shall do so for the most part, when
necessary restoring $R_{ads}$ by dimensional analysis.
 
The $r\mbox{-}\tau$ manifold , $r \in [r_{min},\infty)$ and $\tau \in
[0,\beta)$, at fixed $x^\mu$ can be regarded as the entire 2-d plane
with origin at $r = r_{min}$. At times we will prefer to replace $r$
by the proper distance from $r = r_{min}$,
\be
y(r)   \equiv  \int^r_{r_{min}} 
\frac{dr}{\sqrt{f(r)}} = 
\frac{2}{(d+1)} \log \Big 
[(r/r_{min})^{(d+1)/2} + \sqrt{(r/r_{min})^{d+1} - 1}\Big ] 
\;,
\label{eq:ycoord}
\ee
 or $r(y)=r_{min} 
\cosh^{\frac{2}{d+1}}(\frac{d+1}{2}   y)$.
In terms of $y$, the 
metric becomes
\be
ds^2 = dy^2 +   r^2(y) \tanh^2(\frac{d+1}{2}   y) 
d\tau^2 +   
r^2(y)   dx^\mu dx_\mu \; ,
\ee
with the AdS form recovered in the large y limit where $r/r_{min}
\simeq \exp[ky]$.

\setlength{\unitlength}{1.2 
mm}
\begin{picture}(100,65)
\linethickness{.65mm}
\put(10,35){\vector( 
1,0){80}}
\put(20,30){\line(0,1){10}}
\put(75,30){\line(0,1){10}}
\put 
(22,42){$r = r_{min}$ (IR)}
\put(22,37){$(y = 0)$}
\put(85,42){$r = 
r_c$ (UV)}
\put(85,37){$(y = y_c)$}
\put(10,30){$0$}
\put(85,30){$r 
\rightarrow 
\infty$}
\qbezier(20,35)(20,65)(75,65)
\qbezier(20,35)(20,5)(75,5)
\qbezier(75,65)(70,35)(75,5)
\put(80,60){$\tau$}
\qbezier(75,65)(80,35)(75,5)
\end{picture}

\subsection{Low Energy Effective Action}

Next we introduce a Planck brane at $r = r_c$ replacing the
$r\mbox{-}\tau$ plane by a compact disk, with a $Z_2$ reflection $r
\rightarrow r^2_c/r$. These two disks are then patched together to 
form a $S^2/Z_2$ orbifold of the 2-sphere. In order to see how this
effects the spectrum, we must build a model of the brane and study the
small fluctuations of Einstein's equations subject to the Israel
junction condition.  We model the Planck brane as an infinitely thin
shell by adding a surface term to the action,
\be
S_{eff}  =   - 
\frac{1}{2 \kappa_D} \int_M d^dx d\tau dr \sqrt{-g} \; ( R - 
2
\Lambda )  + \frac{1}{2 \kappa_D} \int_{\partial M} d^dx d\tau 
\sqrt{-q} V_{brane} \; ,
\label{eq:Eaction}
\ee
where $q$ is the determinant of the induced metric. As we will argue 
below, a particular  microscopic
construction can lead to a potential 
of the form
\be
V_{brane} = \lambda_1 + \lambda_2 \exp[- \sigma(x)] 
\; ,
\label{eq:sm}
\ee 
where $g_{\tau\tau} \equiv \exp[2 
\sigma(x)]$.  Unlike the Randall-Sundrum construction, two 
independent parameters, ($\lambda_1,\lambda_2$), are needed because
the cusp due to the $Z_2$ orbifold is not the same in $\tau$ direction
versus the Euclidean $x^\mu$ directions. We shall demonstrate shortly
that in order to have a metrically flat brane on a $Z_2$ orbifold at
$r=r_c$ the parameters in Eq.~(\ref{eq:sm}), must be adjusted to fit
the background metric
\bea
&& \lambda \equiv \frac{d}{d y} \log 
f_0(r_c - \epsilon)
  =\frac{\lambda_1}{2d} \;\; , \nonumber\\ && 
\lambda_\tau \equiv
  \frac{d}{d y} \log f(r_c - \epsilon)= 
\frac{\lambda_1}{2d} + \frac{
  \lambda_2}{2 \sqrt{f}} \; \; 
.
\label{eq:tensions}
\eea
The ratio of $\lambda$ and $\lambda_{\tau}$ fixes the position of the
Planck brane $r = r_c$.

Let us now see if it is possible to construct such an asymmetric brane
from the microphysics that we are allowed to assume.  The brane must
be characterized as an object defined in an intrinsically covariant
fashion, in order that we may place it in the ambient geometry and
consistently couple its fluctuations to the fluctuations of the
metric. This fact places constraints on how we may obtain the
asymmetric brane. Consider a set of $d$-branes with world volume
extending over all the directions $(x^\mu, \tau)$; these branes have
the usual isotropic tension. Now take a collection of $(d-1)$-branes,
and place them perpendicular to the $\tau$ direction.  These branes
have tension only along $x^\mu, \mu=1\dots d$. We let the density of
these branes be uniform along the circle $\tau$. This model leads to a
brane world potential give by Eq. (\ref{eq:sm}).  The combined tension
of this set of branes is clearly asymmetric. The two kinds of branes
are not bound to each other, but since we only consider symmetric
fluctuations, by symmetry both kinds of branes will stay at the same
location $r=r_c$.  We will also study fluctuations that are
independent of the coordinate $\tau$, so the distribution of the 
$d$-branes in the $\tau$ direction will automatically remain uniform.

In this construction we see the difficulties with other ways of
achieving asymmetric tensions.  The basic objects that we have in our
theory are (i) branes and (ii) momentum modes. Suppose we put a set of
1-branes wrapping the direction $\tau$, distributed uniformly in the
directions $x_i$. We do of course want to have vibrations that are
functions of the $x_i$.  But after such a wave passes through this gas
of 1-branes, the 1-branes will no longer be uniformly distributed in
the coordinates $x_i$. Thus their stress tensor will not be mimicked
by a single asymmetric brane; rather we will have to introduce a
density function $\rho(x_i,t)$ to characterize the evolving
distribution of the 1-branes in the directions $x_i$. The same applies
to momentum modes carrying momentum in the direction $\tau$, which we
could naively have considered as another way to generate asymmetric
tensions.

Since we want our physics to be $x^\mu$ dependent (but we can choose
$\tau$ independence), the above construction using $d$-branes and
$(d-1)$-branes appears to be the only simple choice available. However
our subsequent analysis of this brane world leads to instability
signaled by a tachyonic radion mode.  To understand this instability,
we have generalized our model to an effective brane world potential,
\be V_{brane} = \tilde \lambda_1 + \tilde 
\lambda_2
\exp[- \alpha \sigma(x)]\; ,
\label{eq:alpha}
\ee
where $\tilde\lambda_1$ and $\tilde \lambda_2$ are appropriately redefined
to give the flat brane solution. The free parameter $\alpha$ allows
one to adjust the quadratic term about the background, i.e., varying
$V''_{brane}(\sigma_0)$ while holding $V_{brane}(\sigma_0)$ and
$V'_{brane}(\sigma_0)$ fixed in the background metric ($\exp[2
\sigma_0] = f(r)$), thus shifting the radion to positive mass squared
and in turn stabilizing the brane.

\subsection{Definition of Spectral 
Problem}
 
To find the spectrum around this background solutions we must find the
linearized Einstein's equations. More precisely the linearized Euler-Lagrange 
equations of bulk gravity coupled to our low energy effective
brane world action.  Our approach is to write down the EOM in an axial
gauge ($h_{Mr} = 0$ for $M \ne r$) and integrate from the IR horizon
at $r=r_{min}$ to the UV boundary at the Planck brane where
$r=r_c$. Although the actual Minkowski time is parallel to the brane,
we are in a sense treating the radial coordinate as a sort of time
variable. This appears to be the best way to approach the spectral
problem.

We begin by parameterizing the metric in the traditional Kaluza-Klein
form,
\be
ds^2 = \tilde g_{mn} dx^m dx^n + e^{2 \sigma} (d\tau + 
A_m
dx^m)^2 \;  .
\ee
with $x^m = (r,x^1,\cdots x^d)$. As we noted previously in our
glueball analysis~\cite{bmt1,bmt2}, it is natural to classify modes
with respect to the number of $\tau$ indices ( zero for $g_{mn}=
\tilde g_{mn} + \exp[2\sigma]A_n A_m $, one for $g_{m 
\tau} = \exp[2\sigma] A_m$, and two for
$g_{\tau\tau} = \exp[2\sigma]$ respectively). Expanding to linear order in
the fluctuations in axial gauge ( $h_{r M} = 0 $ for $M \ne r$),
\be
ds^2 = f(r)^{-1} (1 + \rho) dr^2 + r^2 (\eta_{\mu \nu} +
 h_{\mu\nu}) dx^\mu dx^\nu + f(r)(1  +S) d\tau^2 + 2 f(r) A_\mu dx^\mu 
d\tau
\; , \label{eq:fluctuations}
\ee
and ignoring $\tau$ dependence (or excitation in the thermal
direction) the trace reversed Einstein equations in the bulk as
follows.\footnote{To be precise about our conventions, the linearized
Einstein equations are written in the form, $ h^M{}_N{}_{;L}{}^{L} +
h_L{}^L{}_;{}^M{}_N - h^M{}_L{}_;{}^N{}_L - h_{NL}{}_;{}^{ML} + 2
(D-2)^{-1} \Lambda h^M{}_N = \Sigma^M{}_N$ where the bulk energy
momentum $\Sigma^M_N = 2 \kappa(T^N{}_M -(D-2)^{-1}g^N{}_M T)$ has
been introduced to set the normalization. The M-index is raised by
the background metric~(\ref{eq:fluctuations}) except when we use
Lorentzian signature $\eta_{\mu\nu}$ (mostly positive) for flat
Minkowski space.}

{\bf Equations of Motion:} We enumerate these equations as 3 second
order wave equations, (tensor, vector and scalar), followed by 3
constraint equations,
\bea
\label{eq:tensor}
 \Sigma_{\mu\nu} =&&  \nabla^2 
h_{\mu\nu}  +
\frac{1}{r^2}[\partial_\mu \partial_\nu ( h + S + 
\rho)
-\partial_\mu \partial^\lambda h_{\lambda\nu}  - 
\partial_\nu
\partial^\lambda h_{\mu\lambda}] \\
&& \qquad \quad + 
[\frac{f}{r} (h' + S'- \rho') - 
 2(d+1)\rho]\eta_{\mu\nu} 
=0\;,\nonumber \\
\label{eq:vector}
 \Sigma^\tau{}_\mu =&& 
\nabla^2  A_\mu  - \frac{1}{r^2}  \partial_\mu
\partial^\lambda 
A_\lambda  + r^2 (\frac{f}{r^2})' A'_\mu = 0 
\;,\\
\label{eq:scalar}
 \Sigma^\tau{}_\tau  =&& \nabla^2  S + 
\frac{ f'}{2 } ( 
 h' + S' - \rho') -  2(d+1) \rho  = 0  \; , 
\\
\label{eq:tconstraint}
\Sigma^r{}_\mu =&&  f \partial^\lambda 
(h'_{\lambda\mu} - \eta_{\lambda\mu}
(h' + S')) +  (\frac{f}{ r}   - 
\frac{ f'}{2 }) \partial_\mu S  +
(\frac{(d-1)f}{r} + \frac{f'}{2}) 
\partial_\mu \rho   = 0  \;, 
\\
\label{eq:vconstraint}
\Sigma^r{}_\tau =&& f \partial^\lambda 
A'_\lambda = 0   \; , \\ 
\label{eq:rr}
\Sigma^r{}_r   =&& f( S'' + 
h'') + \frac{3f'}{2 }  S' + (\frac{2f}{r} + \frac{f'}{2 }) h'    - 
(\frac{d f}{r} +  \frac{f'}{2 }) \rho'
+ 
\frac{1}{r^2}\partial^\lambda\partial_\lambda  \rho  - 2(d+1) \rho 
= 
0 ,  
\eea
where  $h' = \partial_r h$, etc.  We note that the diagonal 
$rr$-equation is  second order in $r$-derivatives; it 
only becomes a first-order constraint in ``trace-unreversed" form, e.g., 
for $\Sigma^r{}_r -
\half (\Sigma^r{}_r + \Sigma^\lambda_\lambda + 
\Sigma^\tau{}_\tau )$. We also note that not all these equations are 
independent due to the Bianchi identity.

It is interesting to point out that these equations have the expected pure 
AdS ($f(r)\rightarrow f_0(r)= r^2$) limits.  It is easy to see that the 3 wave
equations, (\ref{eq:tensor})-(\ref{eq:scalar}), collapse into a
single tensor equation for $h_{ab}$, $a,b = \tau, 1,...,d$, with the
identification $A_\mu \equiv h_{\tau\mu}$ and $S \equiv
h_{\tau\tau}$. Similarly, the
first two constraints, (\ref{eq:tconstraint}) and (\ref{eq:vconstraint}),
after taking into account that $\tau$ derivatives have been dropped, 
also become a single vector equation.  Thus in pure
AdS we have the 3  equations,
\bea
\label{eq:single_tensor}
 \Sigma_{ab} 
=&&   \nabla^2 h_{ab}  +  \frac{1}{r^2}[\partial_a \partial_b ( h + 
\rho) 
-\partial_a \partial^c h_{cb}  - \partial_b \partial^c h_{ac}] 
+  [r (h'- \rho') - 
 2(d+1)\rho]\eta_{ab} =0 \; , 
\nonumber\\
\Sigma^r{}_a =&&  r^2 \partial^c (h'_{ca} - \eta_{ca}
h' ) 
+  d \; r  \partial_a  \rho   = 0  \;,  \nonumber\\
\Sigma^r{}_r 
=&&  r^2 h''  +  3 r h' + (d+1) \rho'
+ 
\frac{1}{r^2}\partial^c\partial_c  \rho  - 2(d+1) \rho \nonumber
= 0 
\; .  
\eea

  Lastly,  we introduce for later convenience a 
differential operator in r, 
 $\nabla^2_r \phi \equiv \sqrt{g}^{-1}(\sqrt{g} f \phi')'$, where $\sqrt{g}= r^d$, 
so that the covariant Laplacian can be expressed as 
$$
\nabla^2 \phi  = \frac{1}{\sqrt{g}}\partial_M (\sqrt{g} g^{MN} \partial_N \phi) =
\nabla^2_r \phi + \frac{1}{r^2} \partial^\lambda \partial_\lambda 
\phi = f \phi'' + (\frac{d \; f}{r} + f') \phi' + \frac{1}{r^2}
\partial^\lambda \partial_\lambda
\phi.
$$

{\bf Boundary Conditions:} The tension on the Planck surface must be
adjusted (or fine tuned) to preserve the Black Hole metric with a
metrically flat ($R^d \times S^1$) brane world at $r = r_c$.  This is
done through the Israel junction conditions, which can also be
 expressed conveniently in Gaussian normal coordinates (or axial gauge
$g_{Mr} = 0$ for $M \ne r$) with the brane fixed at $r = r_c$. The
extrinsic curvature takes the form,
\be
K_{ab}(r_c \pm \epsilon)  =  \frac{1 }{2 
\sqrt{g_{rr}}} \partial_r
g_{ab}(x,r_c
\pm \epsilon) \; ,\label{eq:extrinsic}
\ee
and 
jump condition,
\be
K_{ab}(r_c+\epsilon) - K_{ab}(r_c- \epsilon)
   = 
- 2\kappa_D  \int^{r_c +\epsilon}_{r_c -\epsilon} dr 
(T^{brane}_{ab}
-
\frac{1}{D-2} g_{ab}  T^{brane})  \; ,\label{eq:jump}
\ee
where 
$ g_{ab}$ is the induced metric (namely $a,b$ take on  all 
coordinates
except $r$).

The trace-reversed 
energy-momentum on for our brane model  is
given by
\be
 2\kappa_D(T^{brane}_{ab} - \frac{1}{D-2}   g_{ab} T^{brane} )
  = \left [
\begin{array}{rr}
\lambda_\tau[\sigma] \; g_{\tau \tau} &  0
\\
  0 
& \lambda[\sigma] \;
g_{\mu\nu}
\end{array}
\right ] \; 
\delta(r-r_c)\; ,  \label{eq:em}
\ee
where we define the ``tensions'' in an 
arbitrary background, $\sigma$, by
$\lambda[\sigma] \equiv (V[\sigma] 
+V'[\sigma])/2d$ and $\lambda_\tau[\sigma]
\equiv 
(V[\sigma]-(d-1)V'[\sigma])/2d$.  To zeroth order in the 
background
metric $\exp[2 \sigma_0(r)] = f(r)$, their values are now
fixed by (\ref{eq:extrinsic})-(\ref{eq:em}):  
$\lambda[\sigma_0] = 
\lambda$ and  $\lambda_\tau[\sigma_0] = \lambda_\tau$. These reduce 
to  
 Eq.~(\ref{eq:tensions})  for  $\alpha=1$, as promised.

To 
first order, the Israel junction conditions
provide boundary 
conditions for our EOM at $r = r_c$,
\bea && h'_
{\mu\nu}=[ \frac 
{\lambda }{2} \rho + (\frac{\lambda-\lambda_{\tau}}{2 \sqrt {f} 
})(\frac{1-\alpha}{d})
{S}]     \eta_{\mu\nu} \; ,\nonumber\\
&& S'= 
\frac {\lambda_{\tau}}{2 } \rho +(\frac{\lambda-\lambda_{\tau}}{2 
\sqrt {f}   })(\frac{1-\alpha}{d} + {\alpha}) {S}
\; , \nonumber 
\\
&& A'_\mu = 0 \:.
\label{eq:junctions}
\eea
(We have also imposed $Z_2$ symmetry so that the jump condition is
defined through the radial derivative to the left of the cusp at  $r_{c} -
\epsilon$.)  Note that the simplest model for an asymmetric brane,
Eq.  (\ref{eq:sm}), corresponds to having $\alpha=1$ in which we must
further restrict $A_\mu(r_{c}) = 0$.  To maintain the horizon as a
coordinate singularity also dictates that
\be S(r_{min}) =\rho(r_{min}) \; ,  \ee
and requires that $\rho$, $h_{\mu\nu}$, and $S$ are regular
at $r=r_{min}$~\cite{jev,bmt1,bmt2}.

\newpage
\section{Analysis of Zero Mass  Spectrum}
\label{sec:zeromode}

Randall and Sundrum have observed that the gravitational field in the bulk
spacetime AdS space gives rise to a zero mass graviton.  The graviton
is a domain wall state confined to the Planck brane with an
exponential tail into the bulk when measured in proper distance.  We will see
below that their mechanism for producing brane world  gravity generalizes to our
spacetime background as well.  However in the first Randall-Sundrum model
\cite{RS1} there is another  massless  bulk mode, the ``radion'',
representing the fluctuations in the proper distance separating the
positive and negative tensions branes at the $Z_2$ orbifold planes.
We wish to see if we have a similar radion mode in our model.

To enumerate clearly the physical modes, we choose to completely fix
the gauge. It is natural in our argument to study each wave equation,
as a function of the radial coordinate, by imposing first the boundary condition at
the Black Hole horizon and then integrating out to the Planck brane. A convenient 
gauge for this
analysis begins with our choice of an axial gauge
\be
h_{ra}\rightarrow h'_{ra} = h_{ra} + \xi_{a;r} +  \xi_{r;a} = 0\;,
\qquad\mbox{for} \; a = \tau, 1, \cdots d\; ,
\ee
in a coordinate system with the Planck brane fixed at $r = r_c$ and
the horizon at $r = r_{min}$, independent of the flucutations around
the background. In the above gauge there remains one additional
r-dependent transformation $r \rightarrow r + \xi^r(x,\tau,r)$ with
$$\xi^\mu(x,\tau,r) = - r^2 \partial^\mu \int^r dr
\frac{\xi^r(x,\tau,r)}{r^2  f(r)}   \quad  \mbox{and} \quad
\xi^\tau(x,\tau,r) = - f(r) \partial^\tau \int^r dr
\frac{\xi^r(x,\tau,r)}{  f^2(r)},
$$
as well as $d+1$ r-independent gauge transformations, $x^a \rightarrow x^a +
\xi_0^a(x,\tau)$.  These residual gauge transformations will be used to
eliminate additional unphysical states.  Since the charged KK modes do
not contribute to the zero mass states we may simplify the discussion
by dropping all $\tau$-dependence in what follows, leaving us the
usual gauge transformation $\xi_0^\tau(x)$ for the KK ``photon'',
$h_{\tau \mu} = A_\mu(x)$, and d  r-independent gauge
transformations, $\xi_0^{\mu}(x)$, for our  brane world gravity.

\subsection{Graviton Solution}

We begin by looking for the analog of the Randall-Sundrum solution,
thus obtaining the propagating degrees of freedom for the massless
graviton on the brane.  Our Planck brane spans a d-dimensional Lorentz
space and a compact $\tau$-axis.  We set the gauge field $A_\mu$ to
zero.  By inspection we see from Eq.~(\ref{eq:tensor}) that the
traceless ($\sum_i h^\perp_{ii} = 0$) transverse plane wave state,
$h^\perp_{ij}(r) \exp[ i p_\mu x^\mu ]$, satisfy the equation for  minimally
coupled scalar, 
\be
\quad  \nabla^2_r h^\perp_{ij}  + \frac{m^2}{r^2} h^\perp_{ij} = 0 \; .
\ee
We note that the constant polarization
vectors,  $h^\perp_{ij}(r) = \epsilon_{ij}$, are the solution for a zero-mass 
graviton in  transverse gauge, (with two helicity 
states, $\lambda = \pm 2$, for $d = 4$).

To see that these modes indeed represent only a single graviton in the
boundary theory, we must see that the gauge freedom and gauge
constraints arising from the bulk description agree exactly with the
gauge properties expected from a graviton propagating in the
boundary.  It is useful to
review the standard arguement in flat space by dropping the
r-dependence and the coupling to the additional longitudinal modes
$S,\rho$ in Eq.~(\ref{eq:tensor}). In this case we would have the usual linearized Einstein
equation for the tensor field,
\be
h_{\mu\nu,\lambda}{}^\lambda+h_{,\mu\nu}-h_{\mu\lambda, 
\nu}{}^\lambda-h_{\nu\lambda, \mu}{}^\lambda=0 \; .
\label{eq:flatEH}
\ee
To be explicit, we work in Minkowski space with conventional lightcone axes,
$e^\mu_\pm = (0,...,0,1,\pm 1)$, so that  all vector indices $\mu$,
(e.g., $V_\mu$), are replaced by $i,+,-$, (e.g., $V_i, V_\pm \equiv
e^\mu_\pm V_\mu,$), with $i=1,..,d-2$ and metric $\eta_{+-} = \eta_{-+}
= 2$.  All states are taken to be plane waves with momentum $p_\mu =
(0,...,p,E)$,  $m^2 = -p^2 = - p_+ p_-$ and $p_+ = p + E = 2p$
for $m \rightarrow 0$ so that graviton solution in momentum space is
\be
h_{\mu\nu}=\epsilon_{\mu\nu}e^{i p_+ x^+} \; .
\ee
In Eq.~(\ref{eq:flatEH}), setting $\mu \nu= i+ ,++$, and 
 $+-$,  
we find
\bea
\label{qone}
\epsilon_{i-}=0 \;,  \\
\label{qtwo}
\epsilon-\epsilon_{+-}=0 \; , \\
\label{qthree}
\epsilon_{--}=0 \; ,
\eea
respectively where we have defined $\epsilon =
\epsilon_{\mu\nu}\eta^{\mu \nu}$.
Thus the polarizations found from the bulk
analysis must have the gauge freedom
\be
\epsilon\rightarrow \epsilon_{\mu\nu}+p_\mu\xi_\nu + p_\nu \xi_\nu\; ,
\ee
and be subject to exactly  the above three constraints.  
By the standard arguement~\cite{Feynman}
the physical states in the quotient space are the transverse
graviton polarization $\epsilon_{ij}$ given above.

We now must generalize this argument keeping track of possible
r-dependence and the longitudinal coupling to $S$ and $\rho$. For this
we need to be more precise about the spectrum of the radial equation
for a minimal scalar. This is a Sturm-Liouville eigenvalue problem of
the form,
\be 
- \nabla^2_r \phi_n(r) = \frac{m^2_n}{r^2} \phi_n(r) \; , 
\ee 
with orthonormal condition, $\int^{r_c}_{r_{min}} dr \sqrt{-g} r^{-2}
\phi_n(r) \phi_m(r) = \delta_{nm}$ and a Neumann boundary conditions, 
$\phi'( r_c) =0$ at the Planck brane and regularity at the origin,
$r=r_{min}$.  At $r=r_{min}$, the solution has the form $C_1 + C_2
\log(r- r_{min})$.  Regularity requires that we set $C_2=0$.  The
resulting operator is self-adjoint with a discrete positive
semi-definite spectrum, $m^2_n \ge 0$, and a single null vector, which
is a constant in r: $\phi_0(r) = C$.  Thus the graviton arises as the
unique normalizable state which is present only for finite $r_c$. Our 
earlier tensor glueball calculations~\cite{{bmt1}} with $r_c =
\infty$ exhibited a mass gap simply because this state was excluded
from the Hilbert space due its logarithmically divergent norm. We
postpone to Section 5 a more detailed comment on the effective gravity
theory implied by this solution and its relation to the orginal
Randall-Sundram construction.

\subsection{Longitudinal Modes}

We proceed by solving the full set of  equations, Eqs.\ref{eq:tensor}-\ref{eq:rr}, 
for massless modes. The KK ``photon'' $A_\mu(r)$
has a zero mode but this equation decouples from the rest and in our
effective theory we have set $A_\mu(r_c)=0$ eliminating this mode.
Viewed from the perspective of the little group for a light-like state,
the tensor field $h_{\mu\nu}$, in addition to the graviton, has in general
2(d-2) massless vector modes, ($h_{i \pm }$), and 4 scalars, ($h_{--},
h_{++}$,$h_{+-}$ and $h = \eta^{\mu\nu} h_{\mu\nu} = \sum_i h_{ii} +
h_{+-} $).  These mix with an additional KK scalar, $S$, and the radion,
$\rho$. By finding explicitly all solutions for $m^2 = 0$, we can
settle the issue concerning the existence of additional propagating
scalars, i.e., the radion.  A careful analysis of this sector is a bit
involved. Here we give the outline of the argument.

First, consider first a pair of equations from Eq.~(\ref{eq:tensor}):
\be
-\nabla^2_r h_{i-} = 0 \;\;\; , \;\;\;  -\nabla^2_r h_{i +} = \frac{2 
h_{i -}}{r^2} \; .
\ee
Let us integrate the first equation from $r_{min}$ outwards.  Imposing
regularity at $r = r_{min}$ and satisfying boundary conditions at
$r_c$ gives $h'_{i-} = 0$, i.e., a null state $h_{i-} = constant$.
But the second equation leads to a contradiction if this constant is
non-zero -- we get a logarithmic singularity in $h_{i+}$ at
$r=r_{min}$.  Put suscinctly, the image of our radial operator in the
Hilbert space cannot include the null vector. Thus we conclude that
we must have $h_{i-} = 0$, which reproduces (\ref{qone}), and we then find that
$h_{i+} = {const}$ is also a null vector. This latter constant can be
gauged away to zero by a diffeomorphism $\xi^i\sim x^+$, just as would
be expected for a graviton on the brane.

   Similarly there is a pair of equations for
\be -\nabla^2_r h_{--} = 0
\;\;\; , \;\;\; -\nabla^2_r [ h_{+-} - \frac{2}{d} h] = \frac{2(d-2)
   h_{--}}{d r^2} \; .
\ee
These give the condition $h_{--} = 0$, which reproduces (\ref{qthree}) above,
and the condition that $h_{+-} - \frac{2}{d} h = constant$; this constant 
can again be gauged away by a $r$-independent diffeomorphism.

Next, we note that there are 3 equations for the remaining 4 scalars,
$S,\rho,h_{++}$, and $h$.  This would appear to be an undetermined 
system, but we can perform an $r$-dependent
diffeomorphims to  reduce the radion field to  a single constant,
$\rho(r)
\rightarrow
\rho_0$, representing the fluctuation for the proper distance
separating the Planck brane (at $r=r_c$) from the Black Hole horizon
(at $r= r_{min}$).

Consider the $++$ component of the tensor
equation for $h_{\mu\nu}$
\be
  - \nabla^2_r h_{++} =  - \frac{4}{r^2} (  S + \rho_0 +  h -h_{+-}) 
\; , \label{eq:hpp}
\ee
  and the ``trace-unreversed" form of $\tau \tau$
and the $rr$  equations:
\bea
\nabla^2_r h  - \frac{3 f'}{2}  r^{d-1}  h'  - d(d+1) \rho_0  &=& - 
\frac{1}{r^2}  h_{--} \; ,  \label{eq:einsteintt} \\
d \frac{f}{r}   S'+ ((d-1)\frac{f}{r}  + \frac{f'}{2})  h' - d(d+1) 
\rho_0 &=& -   \frac{1}{r^2} h_{--} \; .
\label{eq:einsteinrr}
\eea
Since we have shown that the function $h_{--}$  is zero, the last two
equations can be integrated rather trivially from $r = r_{min}$, yielding
\be
  S(r) = \frac{f' \rho_0 y }{2\sqrt f}\;, 
\quad
\mbox{and} \quad  h(r)   =  \frac{ d\sqrt f \rho_0 y}{r} + h(r_{min}),
\label{eq:radion2}
\ee
where $y(r) = \int^r_{r_{min}} dr/\sqrt{f(r)}$ is given in
Eq.~(\ref{eq:ycoord}). Note that $S(r_{min})=\rho_0$ and both $h$ and
$S$ are regular at $r_{min}$.  The integration constant $h(r_{min})$
remains to be specified. When we substitute this solution into the
boundary condition at $r = r_c$,
\bea
&&  S'(r_c)= \frac {\lambda_{\tau}}{2 } \rho_0 
+(\frac{\lambda-\lambda_{\tau}}{2 \sqrt {f_c}   })(\frac{1-\alpha}{d} 
+ {\alpha}) {S_c} \; ,\nonumber\\
&& h'(r_c)= \frac {d\lambda }{2} \rho_0 
+(\frac{\lambda-\lambda_{\tau}}{2 \sqrt {f_c}   })({1-\alpha})
{S_c}
\; ,
\label{eq:junctionhs}
\eea
we see that we are forced to set $\rho_0 = 0$, (unless we arbitrarily tune
$\alpha$ to a special value as discussed in the next section). It follows 
from Eq. (\ref{eq:radion2}) that
 we  also get $S=0$ and $h=const$. Returning to
the $h_{++}$ equation and using arguments similar to those above, we 
find that we must have $h-h_{+-}=0$, thus
reproducing  (\ref{qtwo}). (We have previously shown that 
$h_{+-} - \frac{2}{d} h = constant$.
It follows that a single r-independent diffeomorphism can gauge 
both $h_{+-}$ and $h$ to zero.)
Lastly, we also have $h_{++}=constant$, which can 
again be gauged away to zero.

Thus we see that the radion vanishes in the zero mass sector, unless
we tune the parameter $\alpha$ in the above relations to a specific
value; we will discuss such tuning further in the next section. One
might hope that without such tuning we have eliminated the radion and
obtained just a massless graviton on the brane. But as we will see
below the radion actually has {\it negative} $m^2$ at $\alpha=1$,
which is the value of $\alpha$ for our physical construction of the
`asymmetric brane'. Thus the brane construction we started with is
actually unstable. We will explore the nature of the solutions for
arbitrary $\alpha$ in the search for a stable solution; however we do
not know if these other values of $\alpha$ can be reproduced by a
well-defined microscopic matter Lagrangian.

\newpage
\section{Radion Instability}
\label{sec:radion}

To address the question of stability, imagine
expanding the Euclidean action, Eq.~(\ref{eq:Eaction}),
to quadratic order in the fluctuating fields relative to the
AdS Black Hole background.
The quadratic terms will take the form
\be
S_E  =  \frac{\beta}{4 \kappa_D}  \int d^dp \int_1^{r_c} dr \sqrt{g} 
\Big[\phi^\dagger(r)
{\cal L} \phi(r) +  \frac{p^2}{r^2} \phi^\dagger(r) Z\phi(r)  +
\partial_r X_{surface} + X_{brane}\delta(r-r_c) \Big] \; ,
\ee
choosing units so that $k=r_{min}=1$.  ${\cal L}$ and $ Z$ may in
general have Lorentz indices (see examples below).  The general
concept of stability of a classical solution requires that all
fluctuations increase the action so perturbations in the partition
function ($\int dg_{MN} \exp[-S_E]$) are well defined. However it is
well known that off-shell Euclidean Einstein-Hilbert  action is
unbounded and does not even satisfy this stability condition to
quadratic order in flat space.  Consequently we restrict ourselves to
a narrower criterion for on-shell stability where all eigenmodes are
neither tachyonic (negative $m^2$) nor ghost-like (negative norm, i.e.,
negative coefficient for the $p^2$ kinetic term ).  To investigate the
question of on-shell stability, one looks at the condition of
stationarity.  Stationarity of the the surface terms ($X_{surface},
X_{brane}$) is equivalent to imposing the Israel matching condition,
so once we impose these boundary conditions we can drop the surface
terms from the action.  Thus we are led back to the problem of finding
the spectrum by solving the eigenvalue equations, ${\cal
L} \phi(r) = (-p^2/r^2) Z \phi(r)$, for each eigenvector $\phi^{(n)}(r)$
with eigenvalue $p^2 = - m^2_n$. The sign of the norm of an
eigenvector is defined by sign of the kinetic term,
\be 
\int^{r_c}_{1} dr 
\frac{\sqrt{g}}{r^{2}} \; \phi^{\dagger(n)}(r) Z \phi^{(n)}(r) \; , 
\ee 
in the Euclidean action.

\subsection{Radion Mass}

To see if the radion is tachyonic, we begin by examining the
constraint of the two boundary conditions at the Planck brane more
carefully for our potential with a single variable parameter $\alpha$.
Since $S$, $S'$ and $h'$ are homogeneous in $\rho_0$, the trivial
solution $\rho_0=0$ always exists corresponding no massless radion, as
indicated earlier.  When can we have a solution with $\rho_0\neq 0$?
Substituting the solution for $h$ and $S$, (\ref{eq:radion2}), into
the junction conditions, we find remarkably that both equations lead
to a single condition:
\be
\Big(\alpha- 1- \frac {2 d\; f(r_c)}{r_c f'(r_c)}\Big) 
\rho_0=0.
\ee
When  $\alpha$ takes on a critical 
value,
\be
\alpha_{critical}= 1+ \frac {2 d\; f(r_c)}{r_c f'(r_c)}= 
1+ d\;\frac
{\lambda}{\lambda_\tau} ,
\ee
both boundary conditions for $S'$ and $h'$ at $r_c$ are met, with
$\rho_0\neq 0$. Therefore a massless radion can exist by fine tuning
$V_{brane}$.

With $\alpha=\alpha_{critical}$, the remaining fluctuation $ h_{++}$
can next be found by solving Eq.~(\ref{eq:hpp}). 
We obtain the solution 
$$ h'_{++}(r) = \frac{4 }{ r^2 \sqrt {f}}
\;y\;\rho_0 
- \frac{4 \;r_c }{r^3 \sqrt f_c}\; y_c\;\rho_0 \; ,
$$
with $h_{+-}=2 h/d$. Regularity at $r=r_{min}$ has fixed the
integration constant for $h$ so that 
$$h(r_{min})= -\frac
{d(d-1)r_c}{(d-2)\sqrt{f_c} }\;y_c \;\rho_0 \; . $$
The last remaining gauge freedom can be used to specify the
integration constant, thus providing a complete specification for
fluctuations associated with the massless radion.
 
As one smoothly varies $\alpha$ so that $\Delta \alpha \equiv \alpha -
\alpha_{critical}\neq 0 $,  one 
expects this mode to survive with the radion acquiring a mass. For
$\Delta\alpha$ small, this mass can be calculated perturbatively.
Denote $S\simeq S_0 + m^2 S_1$ and $h\simeq h_0 +m^2 h_1,$ where $S_0$
and $h_0$ are solution at $\alpha=\alpha_{critical} $.   Substituting
these into Eqs. (\ref{eq:scalar}) and (\ref{eq:rr}), we find, to first
order in $m^2$, a set of equations for $S_1$ and $h_1$:
\be
 r^2 f(S_1'' + h_1'') + (2 rf + 
\frac{r^2f'}{2 } ) h_1' + \frac{3 r^2f'}{2 }
S_1'  = -\rho_0 
\quad \mbox{and} \quad
    r^2f h_1'' + 2 rf h_1' -  d\; rf\;  S_1' 
=    S_0 -\rho_0 \; .
\ee
By imposing appropriate boundary conditions at both $r=r_{min}$ and $r=r_c$,
these first order perturbations can be obtained. In particular, the
radion mass is fixed to this order.  For $r_c$ large, one obtains a
simple analytic expression
\be
m^2 \simeq  \frac{(d-2)(d+1)}{2 } (\frac{r_{min}}{r_c})^{d-1}\Delta 
\alpha.
\ee
Consequently the system becomes tachyonic when reducing $\alpha$ below
$\alpha_{critical}$ in the direction of our model for the Planck brane
at $\alpha = 1$. We have check this result numerically that finding
for $\alpha < \alpha_{crit}$ a small negative $m^2 < 0$ of order
$0((r_{min}/r_c)^{d-1})$ set by the mass hierarchy. At $\alpha = 1$,
it remains negative, giving $m^2
\simeq -35(r_{min}/r_c)^{d-1}$.

\subsection{Ghost Analysis}

To establish our sign 
conventions
consider first the physical transverse traceless tensor 
field for tensor glueballs
or graviton mode,
$h^\perp_{\mu\nu}(r)$ 
with $p^\lambda h^\perp_{\lambda\nu}(r)= 
 h^\perp_{\mu\lambda}(r) 
p^\lambda = \eta^{\mu\nu}h^\perp_{\mu\nu}(r) = 0$.
Their contribution 
to the action is
\be
S_E  =  \frac{\beta }{4 \kappa_D} \int d^dp dr 
\sqrt{g} \Big[ - h^{\perp *}_{\mu\nu} \nabla^2_r
h^\perp_{\mu\nu} + 
\frac{p^2}{r^2}  h^{\perp *}_{\mu\nu} h^\perp_{\mu\nu} \Big] \; 
.
\ee
The signs are such that all on shell massive modes are neither
tachyonic nor ghost-like. More subtle problem of the decoupling of
null states for the zero-mass graviton has already been discussed
leading to only two physical states to the on-shell graviton to with
$\pm 2$ helicities for d=4.

In the massive scalar sector, we have a coupled problem for four
amplitudes, $h(r,p) =\eta^{\mu\nu}h_{\mu\nu}(r,p)$, $h_T(r,p)
\equiv 
(\eta^{\mu\nu}-\frac{ p^\mu p^\nu }{p^2})h_{\mu\nu}(r,p)$,
$S(r,p)$, 
and $\rho(r,p)$, with
\bea
S_E&=&\frac{\beta }{4 \kappa_D} \int 
d^dp dr \sqrt{g} \Big[ - \frac{d}{d-1}  h^*_{T}\nabla^2_r
h_{T}  + 
h^*_{T}\nabla^2_r
h+ h^*\nabla^2_r
h_{T}  +  S^*\nabla^2_r
h+ 
h^*\nabla^2_r
S    \nonumber\\
 &+&   \frac{(d+1)f}{2} (h^*S'-S^*h') 
+ \frac{p^2}{r^2} \phi^\dagger Z \phi +   {\cal U}(\phi^\dagger,\phi) 
\Big ]
\eea
 The ``kinetic" term is expressed in a matrix form, 
with
\be
\phi^\dagger Z \phi= - \frac{d-2}{d-1}\; h_T^* h_T - 
(h_T^* S+ S^* \rho   +\rho^*h_T+ \mbox{h.c.} ) \; ,
\ee
where the matrix $Z$ defines the norm for these scalar
fluctuations. The remaining non-vanishing terms for $\rho \ne 0$ are
$$
{\cal 
U}(\phi^\dagger,\phi)  =     (\frac{f'}{2}+\frac{(d+1)f}{r})( \rho^* 
h'-h^* \rho') + \frac{d f}{r} (\rho^* S'-S^*\rho ' )
 - \frac{d 
(d+1)}{2} (\rho^*(h_T + S + \rho)+ \mbox{h.c.} ) \; .
$$
Unlike the traceless-transverse tensor modes, the situation is now
more complicated.  Since the Z-matrix has both negative and positive
eigenvalues, the possibility of ghosts is present.  Consequently one
must solve the coupled eigenvalue condition and check the actual
norm, $ \int_1^{r_c}dr r^{d-2}\phi^\dagger Z \phi$, for each mode.

Without the Planck brane, positivity of the norm for on shell physical
states presumably is required if the Maldacena hypothesis relating a
unitary Yang-Mills theory to a ghost free super string is in fact
valid. Nonetheless the quantization and proof of the no-ghost theorem
in Ramond-Ramond backgrounds is far from trivial. Nonetheless for all
the massive glueball states which survive when the Planck brane is
removed are expected to be stable --- neither tachyonic nor
ghost-like.  However, in the presence of the Planck brane new states
do appear.

For the radion we have 
calculated its norm  near the critical point
($\alpha 
\simeq
\alpha_{critical}$) to zeroth order in $m^2$. With $h_T=(\frac 
{d-1}{d})h_0 +0(m^2)$, and, for $r\simeq r_c$ large,
$S_0(r) \simeq 
y(r) \rho_0\;,$ $h_0(r)
\simeq d [ y(r) - (\frac{d-1}{d-2}) y_c ] 
\rho_0\;,
$
it follows  that
\be
 \int_1^{r_c} dr r^{d-2} 
\phi^\dagger Z \phi\simeq
 \frac{1}{(d-2)} 
r_c^{d-1}y_c^2\rho_0^2 >0\;\; ,
\ee
so radion appears not to be a ghost. We have checked numerically that
this result holds for a range of values of $\alpha$ including our
model for the Planck brane at $\alpha = 1$. Therefore, the radion is a
perfectly normal physical mode with a tachyonic mass for our
microscopic construction of the asymmetric brane.

\newpage
\section{ Brane World 
Gravity}
\label{sec:braneworld}

Now let us return to discuss the graviton in our model and compare it
with the Randall-Sundarm constructions.  As noted above the transverse
tensor $h^{\perp}_{\mu\nu}= \epsilon_{\mu\nu}(p) T(r,p) e^{i p x}$,
obeys the same equation as a minimally coupled massless scalar,
\be
\frac{1}{r^d} (r^d f T')'  - 
\frac{p^2}{ r^2} T = 0 \; ,
\ee
with $T'(r_c)=0$.  A minimal scalar field, $\nabla^2 \phi + M^2 \phi =
0$, of mass $M$ has two possible asymptotic forms, $\phi(r) \sim (
r)^{-\Delta}$, or $ \phi(r) \sim ( r)^{\Delta - (d+1)}$, for conformal
dimension $\Delta = \half(d+1) +
\sqrt{\quarter(d+1)^2 + (R_{ads} M)^2}$.  By a
general argument for $AdS^{d+2}/CFT$ duality, only the first
normalizable mode corresponds to an operator in correlation function
for the boundary Yang-Mills theory --- in our present example, $T(r)
\sim ( r)^{-6}$ with  conformal mass $M =0$ corresponding to the 
energy-momentum operator in
the boundary theory.  The second non-normalizable mode is present only
when the boundary theory is deformed by including the operator in the
Yang-Mills Lagrangian. Our new graviton solutions, $T(r) =
\mbox{const} $ with $p^2 = 0$,  apparently corresponds to the second 
possibility where the Planck brane defect provides the deformation in
terms of a UV cut-off so that this state can be normalized.
\begin{figure}[h]
\centerline{\epsfxsize=80mm\epsfysize= 60mm\epsfbox{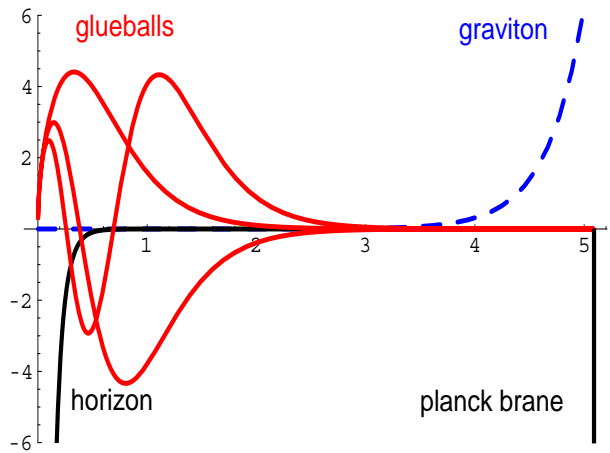}}
\caption{The effective potential $V_{eff}(y)$ (short dashed curve)
with the horizon at $y = 0$ (or $r = 1$) and the delta function Planck
  brane at $y_c = 5.08308$ (or $r_c = 128$). Superimposed profile of
  the graviton wave function (long dashed curve) compared to the
  first three glueball wave functions (solid curve).}
\label{fig:potential}
\end{figure}

For a direct comparison with the Randall-Sundrum solution, it is
instructive to convert to the proper distance (\ref{eq:ycoord}) from
the Black Hole horizon and introduce an integrating factor, $\Psi =
e^{- W(y)} T$, so that the tensor equation has the SUSY form,
\be
[\frac{\partial}{\partial y} - W'(y)][ 
\frac{\partial}{\partial y} +
W'(y)]
\Psi(y)  = \frac{p^2}{ r^2} 
\Psi(r) \; .
\ee
The  ``prepotential''
$$W(y) = - \half  \log(\;r^d 
\sqrt{f(r)}\;) = - \half \log(\;\sinh((d+1)
 y)\;) + 
\mbox{const}$$
is defined in the interval $y \in [0, y_c]$ between the Black Hole
horizon ($y = 0$) and the Planck brane ($y = y_c$).  The effective
potential $V_{eff}(y) = (W')^2 - W''$ is plotted the
Fig~\ref{fig:potential}.  In the limit $r_{min} \rightarrow 0$ (or
large $y \simeq y_c$), we recover the Randall-Sundrum solution, $V(y)
\simeq (d+1) k |y-y_c|$, for pure AdS.

Note the ``volcano'' potential at the Planck brane is a delta-function
at the edge of the disk. In addition, our effective potential has an
attractive double pole at the origin of the disk ($y = 0$) at the
Black Hole horizon ($r = r_{min}$), which accounts for the discrete
glueball or radial KK spectrum. In this representation, the zero-mass
bound-state graviton has wave function, $\Psi = \Psi_0 \exp(- W(y))$
localized close to the Planck brane. Conversely the glueballs are
localized close to the Black Hole horizon with mass on order of
$m_{GB} \sim k^2 r_{min}$.  Note that, with $r_{min} = 0$, there would
be no confinement, and one reverts back to the standard graviton
 equation~\cite{RS2} with no mass gap.

Without the Planck brane, the zero mass solution is no longer a
normalizable state in the tensor spectrum. With the Planck brane, the
graviton appears but the massive states are very slightly shifted by
$\Delta m/m_{GB} = O(r_{min}/r_c)$.  The mass hierarchy between the
low energy glueballs and the Planck scale is therefore a reflection of
the fact that glueballs wave functions are confined to a region close
to $r = r_{min}$ where the graviton is exponentially suppressed as a
function of proper distance, $r_{min}/r_c \simeq \exp[ -k \Delta y]$.

Once we have a graviton and general covariance in the d-dimensional
Minkowski space parallel to our $d+1$ dimensional Planck brane we
expect to find a low energy effective theory for classical brane world
gravity coupled to matter.  To see explicitly how this low energy
theory emerges, we should split the metric into heavy and light modes,
$g_{MN}(r, \tau, x) = \hat g_{MN}(r,x) + h_{MN}(r,\tau, x) $. Assuming
that the graviton is the only zero mode, the {\bf new} background
metric $\hat g$ includes the gravitation fluctuations $\overline
g_{\mu\nu}(x)$ to all orders: $ds^2 =
\frac{1}{f(r)} 
dr^2 + f(r) d\tau^2 + r^2 \overline g_{\mu\nu}(x)dx^\mu dx^\nu \; $.
Expanding the Euclidean action (\ref{eq:Eaction}) in the
$h_{MN}(r,\tau, x)$, we arrive at an effective action. The zeroth
order term gives the dimensionally reduce Einstein-Hilbert action,
\be
M_D^d  \int d^dx \int^\beta_0 d\tau 
\int^{r_c}_{r_{min}} dr \sqrt{-\overline g} \; (kr)^{d-2} \overline 
R(x) 
= M^{d-2}_{Planck} \int d^dx \sqrt{-\overline g} \; \overline R 
\;  .
\ee 
The linear term in $h_{MN}$ should vanish and the quadratic terms
provide the kinetic energy of the massive glueballs coupled
covariantly to gravity.  Higher terms correspond to glueball/graviton
scattering, etc.
 
The strength of the brane world gravity is given by the effective
Planck mass which is related to the transverse volume by
$M^{d-2}_{Planck} = M^d_D \; V_{\perp}$, where
\be 
V_{\perp}= \beta \frac{(k 
r_c)^{d-1} - (kr_{min})^{d-1}}{k(d-1)}  \; .
\ee 
Specifically if we start with 11-d M-theory for the deformed $AdS^7
\times S^4$ background, the above analysis can be summarize by
\be
M^2_{Planck} 
=  \frac{1}{l^9_p} \times (\frac{R}{2})^4 \; Vol(S^4) \times 2 \pi 
R_{11} \times  
\frac{R \; \beta}{4} (\frac{r_c}{R})^4
\; , 
\ee
where the factors reading from left to right respectively come from:
the 11-d Planck scale $M^9_{11} = 1/l^9_p$, the volume of $S^4$, the
length of the 11-th axis and the volume of $r\mbox{-}\tau$ disk
\footnote{Explicitly we use $Vol(S^n) = 2 
\pi^{(n+1)/2}/\Gamma((n+1)/2)$, $R^3 = 8 \pi N l^3_p$, $r_{min} = R^2
\Lambda_{qcd}$, $\alpha'_{qcd} = R^3 l^2_s/r^3_{min}$, $R_{11} = g_s
l_s$, $\beta = 2 \pi R^2/ (3 r_{min})$, and $l^3_p = g_s l^3_s$.}. The
result is
\be
M^2_{Planck}  =   \frac{32  \pi^6 \;  N^2}{9 \alpha'_{qcd}}  \; (r_c/r_{min})^4 \; . 
\ee
It is also worth noting that our radion instability is an extremely
small scale set by the inverse of this hierarchy:
$m^2_{tachyon}/\alpha'_{qcd} =
O(\alpha'_{qcd}/M^2_{Planck})$. Nonetheless such an instability is
still probably too large on a cosmological time scale.
 
If we regard the Black Hole horizon at $r = r_{min}$ as the location
of a ``QCD brane'', this simple expression can be seen as the standard
Randall-Sundrum mass hierarchy~\cite{RS1} between the Planck mass and
the ``QCD'' scale given by the exponential ( $r_{min}/r_c \simeq \exp[
- k \Delta y]$) of the proper distance, $\Delta y$, separating the
Planck brane from the ``QCD brane''.  In addition there is a factor
$N^2$, which sets the scale of the 6 remaining compact dimensions in
units of the fundamental Planck length of M-theory.  In fact we have
found a Randall-Sundrum like mechanism without any explicit
introduction of a second matter brane.  Moreover we have a truly large
$GeV$ scale extra dimensions; however there is no obvious
contradiction with gravity at short distance since the radial KK modes
that give exponential corrections to gravity are simply intermediate
glueball states that are the obvious consequent of QCD corrections to
the graviton propagator.  An alternative way to understand the
hierarchy is to observe that the coupling of glueballs to the graviton
represents an overlap of the graviton wavefunction with the square of
the glueball wave function (see Fig.~\ref{fig:potential}). The small
coupling is now a result of the exponentially small overlap.

\newpage
\section{Conclusion}
\label{sec:conclusion}

It is tempting to speculate that our construct can serve as a simple
model of gravity interacting with non-conformal matter in the confined
phase where both the matter and the graviton are a consequence of
excitations of the same higher dimensional string fields with no need
for external sources. Of course we are aware that this scenario is a
long way from a realistic model that can be rigorously constructed
within the context of non-perturbative string theory.  Even in this
limited context, we have found that our first attempt at a microscopic
theory for the Planck brane, intersecting $d-1$ and $d$ dimensional
branes, actually leads to a tachyonic radion mode.  Future efforts
will explore other approaches to a microscopic theory to see if this
instability can be circumvented. 
 
Finally let us remark on the spectrum in our M-theory approach to the
strong coupling $QCD_4$ glueballs, imagining for now that some
suitable model can be found to stabilize the radion. The glueball
spectrum has been studied in details~\cite{bmt1,bmt2}.  The lowest
mass ``glueball'' states found on the gravitational metric $g_{MN}$
restricted to the $AdS^7$ subspace, obey the inequalities
$$ 
m(0^{++}) \;<\;  m(2^{++}) = m(1^{-+})  =  m(0^{++})\;<\;  m(1^{-+}) 
=  m(0^{-+})   \; .$$
The degeneracy in the spectrum reflect an accidental $O(4)$ symmetry
at strong coupling due to the fact that the compact 11th coordinate
acts like an extra spatial axis. In effect we have a 5-d theory with
tensors, $g_{\tau\tau}$, $g_{\mu\nu}$ and $g_{\mu\tau}$, where $\mu =
1,2,3, 4, 11$. The massive (``glueball'' states) are very slightly
displaced by the presences of a Planck brane due the exponential
hierarchy discussed above.  However, after introducing the Planck
brane, there are a finite number of new ``light'' states. The new
additional states are: (i) A 5-d zero mass tensor graviton with five
physical degrees of freedom which gives rise to Kaluza-Klein gravity
on $R^{(3,1)} \times S^1$ or in 4-d language a $2^{++}$ graviton, a
$1^{--}$ KK photon and a zero mass $0^{++}$ KK dilaton; (ii) a new
radion state, stabilized by some unknown mechanism; and (iii) a new
light 5-d vector again assuming a way is found to introduce a mass
term for the $A_{\mu}$ field coupled to the brane. This decomposes in
4-d into a vector $1^{+-}$/scalar $0^{++}$ multiplet. Note there is
no Planck domain wall state corresponding to the lowest scalar glueball
$0^{++}$ associated with $g_{\tau\tau}$. Instead the radion is an
entirely new degree of freedom due to fluctuations in the finite
proper distances, $\Delta y$, of the Planck brane form the Black Hole
horizon.  Away from strong coupling, the $m(2^{++}) = m(0^{++})$
glueball degeneracy is expected to be broken since the 11-th axis is
distinguished in many ways (being a compact axis on which the membrane
is wrapped) and in fact lattice QCD data also exhibit a significant
splitting.  Consequently a similar fate should lift the mass of the
dilaton.  Once $A_{\mu}$ is reintroduced a standard Higgs mechanism
can lift its mass.  Of course without a viable microscopic model, it
is not possible to be precise, but clearly higher dimensional branes
world models, such as the 5+1 dimensional one we are
considering here, can give very interesting new low mass states.

We have sketched the effective low energy theory on the gravity side
of the AdS/CFT duality. Here the ``glueballs'' and the graviton are
seen as excitations of the same supergravity (or weak coupling string)
modes. This suggests an intriguing possibility for the relationship
between the ``QCD string'' and the ``fundamental string''. In the
confined phase of QCD both strings may merge into a single entity.
One possibility is that when (and if) a phenomenologically
satisfactory vacuum is found for superstring theory, one may have a
phase diagram, which continuously connects quantum gravity to the
lower energy confined QCD phase coupled weakly to low energy
gravitational modes. Of course such a theory would have to have
additional scales in between the Planck brane and the QCD scale for
all of the rest of the standard model, TeV physics and whatever might
``grow'' in the desert.  This need not contradict the alternative dual
description of a weak coupling Yang Mills limit in which gravity is
seen as an additional (semi-classical) background left over from the
superstrings in the IR.  Since the graviton mode is a non-normalizable
state in the standard AdS/CFT dictionary, it presumably corresponds to
adding the Einstein-Hilbert action (at low energy) to the Yang-Mills
Lagrangian.  Although the QCD string in this picture appears simply as
a trick for reformulating the color singlet sector of the confined
phase by String/Gauge duality it is in fact  more directly a
consequence of superstring in extra dimensions. It would be nice to be
able to make this dual view of the QCD string more precise.

\newpage
\def\theequation{A.\arabic{equation}}
\setcounter{equation}{ 
0}
\noindent{\bf Appendix A: Global Radion Analysis in 
Randall-Sundrum Two-Brane Scenario}

Consider a pure $AdS_5$ background, with two fixed branes located at
$r=r_{min}$ and $r=r_c$, ($0<r_{min}<r_c$), and assume
$Z_2$-orbifolding at both branes, (we shall adopt notations as close
to our $AdS/BH$ case as possible.) We are interested in providing a
``global" solution for a massless radion and also in the limit
$r_{min}\rightarrow 0$ where the proper distance between branes
diverges.  Introducing analogous ``reduced" fluctuations, $h_{\mu\nu}$
and $\rho$, the metric, for $r_{min}\leq r\leq r_c$, becomes,
$
ds^2= ({1 +   \rho})\frac {dr^2} {r^2} +  (\eta_{\mu\nu} 
+ h_{\mu\nu}) r^2 dx^\mu dx^\nu
\; .  
$
There are now two sets of junction conditions, at $r=r_{min}$ and
$r=r_c$, where $h_{\mu\nu}'(r_{min}+\epsilon) =\frac
{\rho(r_{min})}{r_{min}}{
\eta_{\mu\nu}} $ and
$h_{\mu\nu}'(r_c-\epsilon) =\frac{\rho(r_c)}{r_c} { \eta_{\mu\nu}} $
respectively.  

Leaving aside the transverse graviton, other massless
modes can again be analyzed as done for the case of AdS/BH using
lightcone variables. There are four
vector fluctuations, $h_{i\pm}$, $i=1,2$, and four scalar
fluctuations, $h$, $h_{++}$, $ h_{--}$, and $h_{+-}$.  One easily
finds that, after global gauge fixings, $h_{i\pm}=0$, $h_{--} =0$, and
$ h= 2 h_{+-}$, with $h$ and $ h_{++}$ remaining to be specified.

We next choose the gauge where $\rho\rightarrow \rho_0$.  
Linearized Einstein equations, 
\be  rh'-4\rho_0 =0\; ,\quad\quad
 \nabla^2_r h_{++} - \frac{ (2 h + 4\rho_0)}{r^2}=0\; ,
\ee
can be solved after enforcing boundary conditions, leading to
\bea
 h(r)&=& 4\rho_0 \Big\{\log (\frac{r}{r_{min}}) -( \frac 
{r_{c}^2}{r^2_c-r_{min}^2})
\log (\frac{r_c}{r_{min}})\Big\}, 
\nn
h'_{++}(r)&=& \frac{4\rho_0}{r^3} \Big\{ \log
(\frac{r}{r_c}) + 
(\frac{r_{min}}{r})^2 (\frac{r_c^2-{r}^2}{r^2_c-{r_{min}}^2}) 
\log
(\frac{r_c}{r_{min}})\Big \}.
\eea
The integration constant for $h_{++}$ can be fixed by the last
r-independent gauge transformation.
It is straight forward to verify that the norm for this mode is
positive.  This massless radion is therefore physical.

Finally, let us find out what happens if one pushes $r_{min}$ to
zero. In this limit, the proper distance between branes $\int_0^{r_c}
\frac
{dr}{r}\rightarrow
\infty$,  thus reducing to a one-brane RS 
scenario~\cite{RS2}. As $r_{min}\rightarrow
0$, both $h$ and 
$h_{++}'$ approach well-defined limits:
$ h(r)\rightarrow 4\rho_0 
\log (\frac{r}{r_c}), $ and
$h'_{++}(r)\rightarrow 
\frac{4\rho_0}{r^3}  \log (\frac{r}{r_c}).$
However, for $h$ and 
$h_{++}$ to be bounded at $r=0$, one
finds that
\be
\rho_0=0\; 
,
\ee
so that the radion decouples, as it should.

\end{document}